\documentstyle[11pt,fullpage]{article}
\def\lsim{\:\raisebox{-0.5ex}{$\stackrel{\textstyle<}{\sim}$}\:}
\def\gsim{\:\raisebox{-0.5ex}{$\stackrel{\textstyle>}{\sim}$}\:}
\def\be{\begin{equation}}
\def\ee{\end{equation}}
\def\ba{\begin{array}{l}}
\def\ea{\end{array}}
\def\eq#1{(\ref{#1})}

\begin{document}

\begin{flushright}
TIFR/TH/94-51
\end{flushright}

\bigskip

\begin{center}
{\Large\bf Like-Sign Dipleton Signature For Gluino Production at LHC
Including Top Quark and Higgs Boson Effects}

\bigskip
\bigskip

Manoranjan Guchait$^a$ and D.P. Roy$^b$\footnote {E-mail:
dproy@theory.tifr.res.in} \\

\bigskip
$^a$ Physics Department \\ Jadavpur University, \\ Calcutta 700 0032, India
\\
\medskip
$^b$ Theoretical Physics Group \\ Tata Institute of Fundamental
Research \\ Bombay 400 005, India

\end{center}

\begin{abstract}

A systematic analysis of the like-sign dipleton signature for gluino
production at LHC is performed in the $R$-conserving minimal
supersymmetric standard model, taking into account the top quark and
Higgs boson effects in the cascade decay.  We consider two
representative values of the gluino mass, 300 and 800 GeV, along with
those of the other SUSY parameters.  While the top quark contribution
is kinematically suppressed for the former case it is very import for
the latter.  Ways of separating the signal from the background are
discussed.  One expects a viable LSD signals upto a gluino mass of
$\sim 800$ (1200) GeV at the low (high) luminosity option of LHC over
practically the full parameter space of MSSM.

\end{abstract}

\section{Introduction}

The large hadron collider (LHC) offers the possibility of squark
$\tilde q$ and gluino $\tilde q$ search right upto the predicted mass
limit of $\sim 1$ TeV [1].  The cannonical search strategy for
these superparticles is based on the missing-$p_T$ signature, which
follows from $R$-parity conservation [2].  The latter implies
that the superparticles are produced in pair and the lightest
superparticle (LSP) resulting from their decay is stable.  It is also
required to be colourless and neutral for cosmological reasons
[3].  In most SUSY models of current interest it is the
lightest neutralino $\chi^0_1$.  The LSP is expected to escape detection
due to its weak interaction with matter very much like the neutrino.
The apparent imbalance of transverse momentum resulting from this
constitutes the missing-$p_T$ signature for superparticle production.

There is a growing realisation in the recent years, however, that an
isolated multilepton and in particular like sign dilepton (LSD)
signature may play an equally important role in superparticle search
[4 - 8].  For the $R$-conserving SUSY model of the
present interest, the main source of lepton in the $\tilde q$ and
$\tilde g$ search at LHC is their cascade decay into the LSP.  They
proceed via the heavier chargino (neutralino) states by emission of a
real or virtual $W(Z)$ boson, which has a significant leptonic
branching fraction -- e.g.
\begin{eqnarray}
\tilde g & \rightarrow & \bar qq'\chi^+_i,\: \chi^+_i \rightarrow
W^+\chi^0_1 \buildrel 0.22 \over \longrightarrow \ell^+\nu \chi^0_1,
\label {1} \\
\tilde g & \rightarrow & \bar qq \chi^0_i,\: \chi^0_i \rightarrow
Z\chi^0_1 \buildrel 0.06 \over \longrightarrow \ell^+\ell^-\chi^0_1,
\label{2}
\end{eqnarray}
where $\ell$ stands for both $e$ and $\mu$.

Recently a systematic analysis of the isolated LSD signature for
gluino pair production at LHC was undertaken in [9] for both
$R$-conserving and $R$-violating SUSY models.  In the $R$-conserving
model, the dilepton final state of interest arises mainly from the
cascade decay \eq{1} of both the gluinos.  After putting in the
leptonic branching fractions of both the $W$ bosons one gets an
overall branching fraction of $\sim 1\%$ for the decay of the gluino
pair into a dilepton final state.  Half of these are expected to be
LSDs since the gluino is a majorana particle.  Despite the small
branching fraction the isolated LSD signature was shown to be viable
for gluino search at LHC because of the small background in this
channel [9].  The main source of background is $t\bar t$
production.  There is a LSD background from the direct leptonic decay
of one $t$ while the other decays into a lepton via $b$.  This is
strongly suppressed by the lepton isolation cut [10].  One also
expects a fake LSD background from the direct leptonic decay of both
$t$ and $\bar t$, where one of the lepton charges is misidentified.

However, ref. [9] did not take into account the effects of top
quark and Higgs bosons in the cascade decay process \eq{1},\eq{2}, which
are important for the resulting LSD signal.  Inclusion of top quark in
the first step of the cascade decay gives [11,12]
\begin{eqnarray}
\tilde g & \rightarrow & \bar t b \chi^+_i + h.c., \label {3} \\
\tilde g & \rightarrow & \bar t t\chi^0_i; \label {4}
\end{eqnarray}
while that of Higgs bosons in the second step gives [12 -- 14]
\begin{eqnarray}
\chi^+_i & \rightarrow & H^+\chi^0_1 \label{5} \\
\chi^0_i & \rightarrow & H^0_k\chi^0_1,\; k = 1-3. \label{6}
\end{eqnarray}

Both the contributions have been studied earlier.  But as far as we
know, there is as yet no systematic analysis of their effects on the
LSD signal at LHC.  The present work is devoted to this exercise.  We
shall estimate the LSD signal arising from the cascade decay of gluino
at LHC using a parton level MC program as in [9], but including the
top quark and Higgs boson effects.  In the process we shall frequently
draw upon the results of earlier works on top quark [11,12] and Higgs
boson [12 - 14] contributions to the cascade decay.  Our emphasis will
be on identifying the main contributers to the leptonic decay of
gluino over different regions of the SUSY parameters, which is
essential for understanding the parametric dependence of the resulting
LSD signal.  It may be noted here that there is a third contribution
to the cascade decay, which was not taken into account in [9] ---
i.e. the loop induced decay
\be
\label{7}
\tilde g \rightarrow g \chi^0_i.
\ee
We shall not consider it here either, since its contribution to
$\tilde g$ decay is very small $(\leq 5\%)$, through out the parameter
space of interest [11,15].  Large values of branching
fractions for \eq{7} reported in [1] (page 627) are
incorrect.  It is worth pointing this out, so that others do not
discover it the hard way we did.

Since the Higgs bosons have negligible couplings to the $e$ and $\mu$
channels, their inclusion in the cascade decay \eq{5},\eq{6} reduces the
leptonic branching fraction of $\tilde g$.  Indeed, we shall see below
that this reduction factor can be quite large ($\sim 2$) over a part
of the parameter space.  Fortunately this is more than offset by the
inclusion of the top quark contribution \eq{3},\eq{4}, which provides an
extra source of $W \rightarrow \ell\nu$.

We have incorporated two other changes in this analysis vis-a-vis ref.
[9].  Firstly the recent MRSD-$'$ parametrisation [16] for gluon
structure function has been used instead of GHR [17].  It
has a considerably steeper gluon, which results in a factor of $\sim
2$ reduction in the signal cross-section from
\be
\label{8}
gg \rightarrow \tilde g \tilde g,
\ee
over the gluino mass range of interest.  We have also cross-checked
this result with the GRV parametrisation [18], which is in
good agreement with the MRSD-$'$ [16].  Secondly the CM energy
has been reduced from 16 to 14 TeV, as currently projected for LHC.
This reduces the signal cross-section by another factor of 2.  Thus
the updates of the gluon parametrisation and the LHC energy reduce
the signal cross-section by a sizeable factor of 4--5 [19].
Fortunately the LSD background from $t\bar t$ is also reduced by a
similar factor, so that the signal/background ratio remains viable
over most of the parameter space of interest.  We have checked that
our signal cross-sections are consistent with those of Baer et al [12]
after taking account of this factor.

In the following section we briefly discuss the cascade decay process
and identify the parameter space of interest.  The gluino mass range
of interest for the LHC can be divided into two parts --- i) the low
mass region $(M_{\tilde g} \sim 300~{\rm GeV})$ in which case the top
quark contributions \eq{3},\eq{4} are kinematically forbidden or
highly suppressed, and ii) the high mass region $(M{\tilde g} \sim
800~{\rm GeV})$ where top quark contributions \eq{3},\eq{4} are
important.  The relevant branching fractions for the cascade decay and
the resulting LSD signals for these two cases are discussed in
sections III and IV.  The main results are summarised in section V.

\section {Cascade Decay Parameters}

The cascade decay formalism has been widely discussed in the
literature [4-6, 20].  We shall only mention the essential points in
order to fix the notation and identify the relevant parameters.  We
shall work within the framework of the minimal supersymmetric standard
model (MSSM), which has the minimum number of parameters.  We shall
generally assume
\be
\label{9}
M_{\tilde g} < M_{\tilde q}
\ee
so that the gluino provides the most important signal for
superparticle production at LHC.  With this assumption the gluino
decay into chargino and neutralino states are insensitive to the
squark mass.  Our numerical results are obtained with a common squark
mass $M_{\tilde q} = M_{\tilde g} + 200~{\rm GeV}$.

There are four neutralino mass eigenstates, which are mixtures of the
four interaction eigenstates, i.e.
\be
\label{10}
\chi^0_i = N_{i1} \tilde B + N_{i2} \tilde W^3 + N_{i3} \tilde H^0_1 +
N_{i4} \tilde H^0_2.
\ee
The masses and compositions of the neutralinos are obtained by
diagonalising the mass matrix
\be
\label{11}
M_N = \pmatrix{M_1 & 0 & -M_Z\sin\theta_W \cos\beta
& M_Z\sin\theta_W\sin\beta \cr
0 & M_2 & M_Z\cos\theta_W \cos\beta & -H_Z \cos\theta_W \sin\beta \cr
-M_Z\sin\theta_W\cos\beta & M_Z\cos\theta_W\cos\beta & 0 & -\mu \cr
M_Z\sin\theta_W\sin\beta & -M_Z\cos\theta_W\sin\beta & \mu & 0}
\ee
where $\mu$ is the supersymmetric Higgsino mass parameter and
$\tan\beta$ is the ratio of the two Higgs vacuum expectation values.
$M_1$ and $M_2$ are the soft masses for $bino \: \tilde B$ and
$wino \: \tilde W$ respectively, which are related to the gluino mass
in MSSM, i.e.
\begin{eqnarray}
\label{12}
M_2 &=& {\alpha \over \sin^2\theta_W\alpha_s}\: M_{\tilde g} \simeq 0.3
M_{\tilde g},\nonumber \\
M_1 & = & {5 \over 3}\: \tan^2\theta_W M_2 \simeq 0.5\: M_2.
\end{eqnarray}
We have followed the analytical prescription of [21] for diagonalising
this mass matrix, but cross-checked our results with the numerical
diagonalisation program ${\rm EISCH1.FOR}$ as well as the published
results of [1,4,22].

The chargino mass matrix
\be
\label{13}
M_C = \pmatrix{M_2 & \sqrt 2\: m_W\sin\beta \cr \sqrt 2\: M_W\cos\beta
& \mu}
\ee
is diagonalized via the biunitary transformation
\be
\label{14}
U\: M_c\: V^{-1}
\ee
to obtain the mass eigenvalues.  The corresponding chargino
eigenstates are
\begin{eqnarray}
\chi^\pm_{iL} & = & V_{i1}\tilde W^\pm_L + V_{i2} \tilde H^\pm_L,
\nonumber \\
\label{15}
\chi^\pm_{iR} & = & U_{i1} \tilde W^\pm_R + U_{i2} \tilde H^\pm_R,
\end{eqnarray}
where $L$ and $R$ refer to left and right chirality states.  We shall
use the real orthogonal representation of the unitary matrices $U, V$
and $N$.  The chargino and neutralino states will be labelled in
increasing order of mass, with $\chi^0_1$ representing the LSP.

Thus the chargino and neutralino masses and compositions are specified
in terms of the three parameters --- i) $M_{\tilde g}$, ii) $\mu$ and
iii) $\tan \beta$ --- which in turn determine the cascade decay
processes \eq{1},\eq{2}.

\begin{enumerate}

\item[{i)}] The gluino mass range of interest at LHC is 200
GeV to $\sim 1$ TeV.  We shall choose two representative values
\be
\label{16}
M_{\tilde g} = 300 \; {\rm and} \; 800\: {\rm GeV}
\ee
corresponding to a relatively light and heavy gluino.  As we shall see
below, the relevant cascade decay processes for the two cases are very
different.

\item[{ii)}] There are two distinct regions in the $\mu$ parameter
space corresponding to $|\mu| > M_2$ and $|\mu| < M_2$.  It is
intuitively clear from \eq{11} and \eq{13} that the lighter chargino
$(\chi^\pm_1)$ and neutralinos $(\chi^0_{1,2})$ are gaugino dominated
in the former case and Higgsino dominated in the latter.  For the
above range of $M_{\tilde g}$, the region
\be
\label{17}
-40\: {\rm GeV} \lsim \mu \lsim 80\: {\rm GeV}
\ee
is excluded by the LEP data [22].  For it corresponds to a Higgsino
dominated $\chi^0_1$ with ${\rm mass } < M_Z/2$, which would show up
in $Z$ decay.  We shall take two pairs of values consistent with the
LEP limits \eq{17},
\be
\label{18}
\mu = \pm 4 M_W\; {\rm and} \; \pm M_W,
\ee
which represent the two regions mentioned above.

\item[{iii)}] The results are rather insensitive to $\tan \beta$ over
the range allowed by MSSM,
\be
\label{19}
1 < \tan \beta < m_t/m_b (\simeq 40),
\ee
except for some Higgs contributions as discussed below.
We shall choose $\tan \beta =2\; {\rm and}\; 10$ as two representative
values.  The current lower mass bounds of $H^0_2$ seems to
disfavour $\tan\beta \simeq 1$ (see eqs. 20 and 24 below), although it
cannot be ruled out in view of the large radiative correction [23].

\end{enumerate}

Inclusion of top quark in the cascade decay \eq{3},\eq{4} requires the
knowledge of $m_t$.  We shall use the value, $m_t = 175~{\rm GeV}$,
suggested by the recent CDF data [24].  Finally, the inclusion of
Higgs bosons in the cascade decay \eq{5},\eq{6} brings in one more
parameter, which can be taken as the charged Higgs boson mass.  Then
the neutral Higgs boson masses are given by the MSSM mass relation (at
tree level) [13,20]
\be
\begin{array}{l}
\label{20}
M^2_{H^0_3} = M^2_{H^\pm} - M^2_W, \\
M^2_{H^0_1, H^0_2} = {1 \over 2}\: \left[M^2_{H^0_3} + M^2_Z \pm
\left\{ \left(M^2_{H^0_3} + M^2_Z\right)^2 - \left(2M_Z M_{H^0_3} \cos
2\beta\right)^2 \right\}^{1\over2} \right]
\end{array}
\ee
which imply
\be
\label{21}
M_{H\pm} > M_W (80~{\rm GeV}),\; M_{H^0_2} < M_Z (91~{\rm GeV}).
\ee
We shall consider 2 extreme cases
\begin{eqnarray}
\label{22}
M_{H^\pm} & = & 500 \Longrightarrow M_{H^0_3} = 494,\; M_{H^0_1} = 499
(494),\; M_{H^0_2} = 54 (89), \nonumber \\
M_{H^\pm} & = & 100 \Longrightarrow M_{H^0_3} = 60, \; M_{H^0_1} = 104 (92),\;
M_{H^0_2} = 31 (58),
\end{eqnarray}
for $\tan \beta = 2 (10)$, where all the masses are in GeV.  Thus to a
first approximation
\be
\begin{array}{l}
\label{23}
M_{H^\pm} \sim 500~{\rm GeV}~
\Longrightarrow M_{H^0_1}, M_{H^0_3} \sim 500~{\rm GeV}, M_{H^0_2}
\lsim 100~{\rm GeV}, \\
M_{H^\pm} \sim 100~{\rm GeV}~ \Longrightarrow
M_{H^0_1}, M_{H^0_3}, M_{H^0_2} \lsim 100~{\rm GeV},
\end{array}
\ee
which will be adequate for our purpose.
In the first case only $H^0_2$ will participate in the cascade decay
while in the second case all the Higgs bosons will participate in it.
The LEP mass limits for these Higgs bosons are [23]
\be
\label{24}
M_{H^\pm} > 41.7~ {\rm GeV}, M_{H^0_3} > 22~ {\rm GeV}, M_{H^0_2} >
44~{\rm GeV}.
\ee

\section{Signature for Low Mass Gluino $(M_{\tilde g} = 300\:{\rm
GeV})$}

In this case the top quark contributions \eq{3},\eq{4} are
kinematically forbidden (or strongly suppressed).  Thus only the
light quarks participate in the $\tilde g$ decay into $\chi^\pm_i,
\chi^0_i$ as shown in the 1st steps of \eq{1},\eq{2}.  Consequently
one can neglect the interference terms between the right and left
handed squark exchange amplitudes.  More importantly one can also
neglect the Yukawa couplings associated with the Higgsino compontents
of $\chi^\pm_i$ and $\chi^0_i$, so that the $\tilde g$ decay into
these states are only governed by their gaugino components in
\eq{10},\eq{15}.  The relevant decay amplitudes as well the
compostions of the $\chi^\pm_i$ and $\chi^0_i$ states can be found in
[9].  We show the resulting branching fractions along with the
$\chi^\pm_i$ and $\chi^0_i$ masses in Table I for convenience.  The
dominant decay channels of gluino are seen to be the lighter chargino
$(\chi^\pm_1)$ and neutralino $(\chi^0_{1,2})$ states, which are
gaugino dominated.  This is true not only at $\mu = \pm 4~M _W$, but
remains approximately valid even at $\mu = - M_W$, which is close to
boundary of the allowed parameter space \eq{17}.  The reason of course
is that the condition $|\mu| > M_2 (\simeq 0.3~M_{\tilde g})$ holds
practically throughout the allowed parameter space for $M_{\tilde g}
\simeq 300\: {\rm GeV}$.  Note that $\mu = M_W$ is already disallowed
by the LEP data as indicated by the corresponding $\chi^0_1$ and
$\chi^\pm_1$ masses.

Coming to the 2nd step of the cascade decay, we see that the only
decays of $\chi^\pm_1$ and $\chi^0_2$ kinematically allowed are
three-body decays into the LSP $(\chi^0_1)$ via virtual gague $(W,Z)$
or Higgs $(H^\pm,\: H^0_k)$ bosons [25].  We shall give the formulae
for the three-body decay widths since they are not readily available
in the literature.  They have been derived using the Feynman rules of
[20].
\be
\begin{array}{l}
\label{25}
\Gamma_{\chi^\pm_i \buildrel W \over \rightarrow \chi^0_j \bar ff'}
=  {9g^4 \over (8\pi M_i)^3} \int ds\: \lambda^{1\over2} \left(M^2_i,
M^2_j, s\right) \\
\hbox{~~~~~~~~~~~~~} {.\left[{1 \over 3} \left(G^2_L + G^2_R\right)
\left\{\left(M^2_i - M^2_j\right)^2 + s\left(M^2_i + M^2_j -
2s\right)\right\} - 4G_LG_R \epsilon_i\epsilon_j M_i M_j s\right]
\over (s - M^2_W)^2}  \\
G_L  =  N_{j2} V_{i1} - {1 \over \sqrt{2}}\: N_{j4} V_{i2},
\\
G_R  =  N_{j2} U_{i1} + {1 \over \sqrt 2}\: N_{j3} U_{i2},
\end{array}
\ee
where $\epsilon_i, \epsilon_j$ represent the signs of the $\chi^\pm_i,
\chi^0_j$ masses and sum over $f$ and $f'$ is understood.  As usual
\be
\label{26}
\lambda \left(M^2_i, M^2_j, s\right)  =  \left(M^2_i + M^2_j -
s\right)^2 - 4M^2_i M^2_j.
\ee
\be
\begin{array}{l}
\label{27}
\Gamma_{\chi^\pm_i \buildrel H^\pm \over \rightarrow \chi^0_j \bar
ff'}  =  {g^4 \over (8\pi M_i)^3}\; {m^2_\tau \tan^2\beta \over 2M^2_W}
\int ds\: \lambda^{1\over 2} \left(M^2_i, M^2_j, s\right)  \\
\hbox{~~~~~~~~~~~~~} {. \left[\left(F^2_L + F^2_R\right)\: \left(M^2_i + M^2_j
- s\right) + 4 F_LF_R \epsilon_i\epsilon_j M_i M_j\right] s \over
\left(s - M^2_{H^\pm}\right)^2}  \\
F_L  =  \cos \beta \left[N_{j4} V_{i1} + {1 \over {\sqrt 2}}
\left(N_{j2} + N_{j1} \tan \theta_W\right) \: V_{i2}\right],
\\
F_R  =  \sin \beta \left[N_{j\beta} U_{i1} - {1 \over \sqrt 2}\:
\left(N_{j2} + N_{j1} \tan \theta_W\right) U_{i2}\right],
\end{array}
\ee
where the factor in front of the integral comes from $H^\pm$ coupling
to the $\tau \nu$ channel, which dominates its decay for $\tan\beta >
1$.
\be
\begin{array}{l}
\label{28}
\Gamma_{\chi^0_i \buildrel Z \over \rightarrow \chi^0_j \bar ff}  =
{4g^4 \sum_f 2\left(g_V^{f^2} + g_A^{f^2}\right)\: G'^2 \over
\cos^4\theta_W \left(8 \pi M_i\right)^3} \int ds \lambda^{1 \over 2}
\left(M^2_i, M^2_j, s\right) \\
\hbox{~~~~~~~} {. \left[ {1 \over 3} \left\{\left(M^2_i -
M^2_j\right)^2 + s\left(M^2_i + M^2_j - 2s\right)\right\} +
2\epsilon_i \epsilon_j M_iM_j s\right] \over (s - M^2_Z)^2} \\
G'  =  {1 \over 2}\: (N_{i3} N_{j3} - N_{i4} N_{j4}), \\
\sum_f 2(g^{f^2}_V + g^{f^2}_A) = \sum_f \left(T^{f^2}_3 -
\sin^2 \theta_W T_3^f Q^f + 2 \sin^4 \theta_W Q^2_f\right) = 3.6,
\end{array}
\ee

where the summation runs over all the leptons and quarks upto $b$, and
$\sin^2\theta_W = 0.23$ [23].
\begin{eqnarray}
\label{29}
\Gamma_{\chi^0_i \buildrel H^0_k \over \rightarrow \chi^0_j \bar ff} &
= & {3 g^4m^2_b a^2_k F^2_{ijk} \over M^2_W \cos^2\beta (8\pi M^3_i)}
\int ds \lambda^{1\over 2} \left(M_i^2, M^2_j,s\right) \nonumber \\
&& \hbox{~~~} {. \left[\left(M_i + \epsilon_i\epsilon_j \eta_k
M_j\right)^2 - s \right] \: s \over \left(s - M^2_{H^0_k}\right)^2},
\nonumber \\
F_{ijk} & = & {1\over 2}\: e_k\left[N_{i3} N_{j2} + N_{j3} N_{i2} -
\tan \theta_W \left(N_{i3} N_{j1} + N_{j3} N_{i1}\right)\right]
\nonumber \\
& + & {1 \over 2}\: f_k\: \left[N_{i4}N_{j2} + N_{j4} N_{i2} - \tan
\theta_W \left(N_{i4} N_{j1} + N_{j4} N_{i1}\right)\right], \nonumber
\\
\eta_{1,2,3} & = & 1,1, -1 \nonumber \\
a_{1,2,3} & = & \cos \alpha,\: \sin \alpha \: \sin\beta \nonumber \\
e_{1,2,3} & = & -\cos \alpha,\: \sin \alpha, \: \sin \beta \nonumber \\
f_{1,2,3} & = & \sin \alpha,\: \cos \alpha,\: \cos\beta
\end{eqnarray}
where
\be
\label{30}
\tan 2\alpha = \tan 2\beta \: \left({M^2_{H^0_1} + M^2_{H^0_3} \over
M^2_{H^0_3} - M^2_Z}\right).
\ee
An alternative but equivalent expression for $F_{ijk}$ is given in
[13].

Let us compare the $\chi^\pm_1 \rightarrow \chi^0_1$ decay widths via
virtual $W$ and $H^\pm$ given in \eq{25} and \eq{27} respectively.
The factors in front of the itegrals came from the decay vertices of
the virtual $W$ and $H^\pm$ bosons.  The latter is relatively
suppressed by a factor $\sim 3 \times 10^{-5} \tan^2\beta$, which is
$\ll 1$ throughout the $\tan \beta$ range \eq{19} of interest. Besides
there is a larger propagator suppression factor for \eq{27} relative
to \eq{25} in view of the mass inequality \eq{21}.  The remaining
factors are expected to be comparable, since $G_{L,R} \sim F_{L,R}$;
they represent $W$ and $H^\pm$ couplings to $\chi_1^\pm \chi^0_1$ and
are each suppressed by an offdiagonal element of the composition
matrices.  Thus one can safely neglect the Higgs boson contribution to
the $\chi^\pm_1 \rightarrow \chi^0_1$ decay.  Hence the cascade decay
\be
\label{31}
\tilde g \rightarrow \bar qq'\chi^\pm_1,\; \chi^\pm_1 \buildrel 1 \over
\rightarrow W\chi^0_1 \buildrel 0.22 \over \rightarrow \ell\nu\chi^0_1
\ee
is expected to provide a leptonic branching fraction of
\be
\label{32}
0.22B_{\tilde g \rightarrow \bar qq' \chi^\pm_1} \sim 0.1
\ee
over most of the parameter space (see Table I) [26].

Similar comparison between the $\chi^0_2 \rightarrow \chi^0_1$ widths
via \eq{28} and \eq{29} shows that the factors in front of the
integrals, coming from the decay vertices of the virtual $Z$ and
$H^0_k$, are already comparable for $\tan \beta = 10$.  Moreover the
$\chi^0_2 \chi^0_1$ coupling of $Z$ is smaller than those of $H^0_k -
{\rm i.e.}\: G' < F_{21 k}$.  For $G'$ is suppressed by two offdigonal
elements of the composition matrix, since $Z$ couples to a pair of
neutralinos only through their Higgsino components.  Finally the
propagator suppression for $Z$ is larger than that of $H^0_2$ because
of \eq{21}.  Thus the virtual Higgs contribution to the $\chi^0_2
\rightarrow \chi^0_1$ decay is expected to dominate over the
$Z$ contribution for $\tan \beta \geq 10$.  Hence the leptonic decay
mode of $\tilde g$ via $\chi^0_2$
\be
\label{33}
\tilde g \rightarrow \bar q q \chi^0_2,\; \chi^0_2 \rightarrow
Z\chi^0_1 \buildrel .06 \over \rightarrow \ell^+\ell^- \chi^0_1
\ee
is suppressed over a large part of the parameter space.  We shall
neglect this mode altogether in estimating the LSD signal below.  It
may be added here that, without the Higgs contribution, this process
would effectively add $\sim 40\%$ to the leptonic branching fraction
of \eq{31},\eq{32} and hence double the resulting LSD signal [9].  Of
course its contribution to the LSD signal comes from the 3 and 4
lepton final state, while the exclusive LSD state comes only from
the decay mode \eq{31} of the gluino pair.

Fig.1 shows the isolated LSD signal from \eq{31} against the $p_T$ of
the softer lepton for $\mu = \pm 4M_W$ and $-M_W$.  In the last case
there is also a modest contribution via the $\chi^\pm_2$ state.  An
isolation cut of
\be
\label{34}
E^{Ac}_T < 10~ {\rm GeV}
\ee
has been applied on both the leptons, where $E^{Ac}_T$ is the
transverse energy accompanying a lepton within a cone of $\Delta R =
(\Delta\phi^2 + \Delta \eta^2)^{1/2} = 0.4$.  A nominal rapdity cut of
$|n_\ell| < 3$ has also been applied on the leptons.  The signals is
seen to be remarkably insensitive to the choice of $\mu$ as well as
$\tan \beta$.  As per \eq{31},\eq{32} about 1\% of the gluino pair
decay into a dilepton final state, so that the LSD signal occurs at
the level of $\sim 1/2$\% of the $\tilde g\tilde g$ cross-section.

Fig. 1 also shows the two backgrounds from $t\bar t$ production
mentioned above --- i.e. the LSD background from
\be
\label{35}
t \rightarrow b\ell^+\nu,\; \bar t \rightarrow \bar b \rightarrow \bar
c \ell^+\nu,
\ee
as well as a fake LSD background from
\be
\label{36}
t \rightarrow b\ell^+\nu,\; \bar t \rightarrow \bar b\ell^- \bar \nu,
\ee
where one of the lepton charges is misidentified.  The fake LSD
background has been set at the level of 1\% of the cross-section for
\eq{36}.  The isolation cut effectively suppresses the LSD background
\eq{35} for $p_{T^2} \gsim 50~{\rm GeV}$ [10], but not the fake
background from \eq{36}.  However, there is a large amount of
missing-$p_T$ acompanying the LSD signal, thanks to the $\nu$ and
$\chi^0_1$ in \eq{31}.  This can be exploited to separate the signal
from the backgrounds, as we shall see later in Fig. 4 [27].

\section {Signature for High Mass Gluino $(M_{\tilde g} = 800\:{\rm
GeV})$}

In this case one has to include the top quark contributions
\eq{3},\eq{4} along with the light quark contributions to gluino decay
shown in the first steps of \eq{1},\eq{2}.  The latter contributions
are calculated in the same way as before.  The results are also very
similar execpt for one case.  At $\mu = -M_W$, the magnitude of $\mu$
is now significantly lower than $M_2$, so that the lighter chargino
and neutralino states are Higgsino dominated.  Consequently gluino
decays preferencially into the heavier chargino $(\chi^\pm_2)$ and
neutralino $(\chi^0_{3,4})$ states, which are gaugino dominated.  The
branching fractions are shown in Table II for $\mu = -M_W$ and $4M_W$.
The results for $\mu = -4M_W$ are similar to those of $4M_W$ and hence
not shown.  Instead we include the point $\mu = 6M_W$, which will be
relevant for the top quark contributions discussed below.

For the gluino decay processes \eq{3},\eq{4} involving the top quark
one has to include the interference terms between the right and left
handed squark exchanges as well as the Yukawa couplings associated
with the Higgsino components of $\chi^\pm_i$ and $\chi^0_i$.
Consequently the squared matrix elements for \eq{3} and \eq{4} are
very long.  These are given in ref. [11].  We have used them in
calculating the decay rates for \eq{3} and \eq{4}.  The resulting
branching fractions are shown in Table II.

Here the Higgsino components of $\chi^\pm_i$ and $\chi^0_i$ play a
very important role in determining the gluino branching fractions.  At
$\mu = -M_W$, where the Higgsino dominated states $\chi^\pm_1$ and
$\chi^0_{1,2}$ are kinematically favoured, they account for the
largest branching fractions of gluino unlike the light quark
contribution.  At $\mu = 4\:M_W$, the Higgsino dominated states
$\chi^\pm_2$ and $\chi^0_{3,4}$ have an equal share of the branching
fractions in spite of being kinematically disfavoured.  Only at $\mu =
6\:M_W$, these Higgsino dominated states become kinematically
inaccessible, so that the gluino decays only into the gaugino
dominated ones.  It is interesting to note that the net top quark
contribution to the gluino decay is as large as 70\% at $\mu = -M_W$,
going down to 40\% at $4\:M_W$ and 20\% at $6\:M_W$.  The last value
holds for the $\mu > 6\: M_W$ region as well.  Its excess share at
lower values of $\mu$ arises from the Yukawa couplings of charginos
and neutralinos associated with the large top quark mass.  It is
important for the resulting LSD signal as we shall see below.

We shall be interested in the leptonic branching fractions of top,
\be
\label{37}
t \buildrel 1 \over \rightarrow bW \buildrel 0.22 \over \rightarrow
b\ell\nu,
\ee
as well as the chargino and neutralino states of Table II.  Clearly
\be
\label{38}
\chi^\pm_1 \buildrel 1 \over \rightarrow W\chi^0_1 \buildrel 0.22
\over \rightarrow \ell \nu\chi^0_1
\ee
whether the $W$ is real or virtual, since the competing $H^\pm$ will
be always virtual and hence suppressed for the reasons discussed
earlier. For the same reason we expect the
\be
\label{39}
\chi^0_2 \rightarrow H^0_2 \chi^0_1
\ee
to dominate over a large part of the parameter space and hence not be
of interest for the leptonic decay.

We have to also consider the leptonic decays of the heavier chargino
and neutralino states here.  It is clear from their masses that they
undergo two-body decay emitting real $W,Z$ or Higgs bosons.  These
two-body decays have been widely discussed in the literature [12-14].
The $\chi^0_{3,4}$ decays are Higgs dominated over large parts of the
parameter spaces; and besides their contributions to gluino decay are
relatively small.  Therefore we shall concentrate on the $\chi^\pm_2$
decay.  Moreover we shall use the asymptotic values of its branching
fractions given in [13], for large $M_{1,2}$ or $|\mu|$, since they are
simple and accurate enough for our purpose.  For $|\mu| > M_{1,2}$,
which hols for the point $\mu = 4\:M_W$, the relative branching
fractions of $\chi^\pm_2$ into the following channels
\be
\label{40}
\chi^\pm_2 \rightarrow \chi^0_1 W,\: \chi^0_1 H^\pm, \: \chi^0_2 W,\:
\chi^0_2 H^\pm,\: \chi^\pm_1 Z,\: \chi^\pm_1 H^0_1,\: \chi^\pm_1
H^0_2, \chi^\pm_1 H^0_3
\ee
are $\tan^2\theta_W (\simeq 1/3)$ for the first two and 1 for the
rest.  Recall from \eq{23} that all the five Higgs channels are
kinematically accessible for $M_{H^\pm} \sim 100~{\rm GeV}$, while
only the $H^0_2$ channel is accessible for $M_{H^\pm} \sim 500~{\rm
GeV}$.  Consequently one has the following branching fractions for
$\chi^\pm_2$ for $M_{H^\pm} \sim 500 (100)~{\rm GeV}$:
\be
\begin{array}{l}
\label{41}
\chi^\pm_2 \buildrel 0.1(.05) \over \rightarrow \chi^0_1 W \buildrel
.22 \over \rightarrow \chi^0_1 \ell\nu, \\
\chi^\pm_2 \buildrel 0.3(.15) \over \rightarrow \chi^0_2 W \buildrel
.22 \over \rightarrow \chi^0_2 \ell \nu, \\
\chi^\pm_2 \buildrel 0.3(.15) \over \rightarrow \chi^\pm_1 Z \buildrel
0.6 \over \rightarrow \chi^\pm_1 \ell^+\ell^-.
\end{array}
\ee
For $|\mu| < M_{1,2}$, correspoding to the point $\mu = -M_W$, the
relative branching fractions of all the channels in \eq{40} are equal.
Consequently one get
\be
\begin{array}{l}
\label{42}
\chi^\pm_2 \: \buildrel .25(.12) \over \longrightarrow\: \chi^0_1W \:
\buildrel .22 \over \longrightarrow\: \chi^0_1 \ell\nu, \\
\chi^\pm_2\: \buildrel .25(.12) \over \longrightarrow\: \chi^0_2W \:
\buildrel .22 \over \longrightarrow\: \chi^0_2 \ell \nu, \\
\chi^\pm_2\: \buildrel .25(.12) \over \longrightarrow\: \chi^\pm_1 Z\:
\buildrel .06 \over \longrightarrow\: \chi^\pm_1 \ell^+\ell^-,
\end{array}
\ee
for $M_{H^\pm} \sim 500 (100)~{\rm GeV}$.  We shall not consider the
lepton from the $\chi^\pm_1$ decay in the last line as it would be
further degraded in $p_T$.

The LSD signal comes from the leptonic decay of each gluino arising
from any of the above processes.  For $\mu = -M_W$, the major
contributions are from
\be
\label{43}
\tilde g \rightarrow \bar q q' \chi^\pm_2,\: \bar tb\chi^+_1 + h.c.,\:
\bar t b\chi_2^+ + h.c., \: \bar t t\chi^0_i
\ee
followed by the leptonic decays of \eq{37}, \eq{38} or \eq{42}.  The
LSD signal so calculated includes the contributions from the 3 and 4
lepton states; but the size of these contributions is relatively
small.  The 3 lepton contribution accounts for a little under a
quarter of the signal while the 4 lepton contribution is negligible.
The resulting LSD signal is shown in Fig. 2 for $\tan\beta = 2$ and
10.  In each case the signal is shown for the two extreme choices of
$M_{H^\pm} = 500$ and 100 GeV.  It is seen to be insensitive to either
of these parameters.  Note that the light quark contribution to the
leptonic branching fraction of gluino comes mainly from the
$\chi^\pm_2$ decay \eq{42}, where the reduction factor from the Higgs
boson effect can be as large as 3/8.  This is more than offset,
however, by the top quark contribution so that the net LSD signal is
large as well as insenstive to the Higgs boson effect.  The size of
the LSD signal is at the level of $\sim 4\%$ of the $\tilde g\tilde
g$ cross-section; i.e. an order of magnitude larger than the case
discussed earlier.  The major part of this signal comes from the
leptonic decay of the gluino pair via top.  It is a characteristic
feature of the majorana nature of the gluino, however, that the pair
of top quarks and the resulting leptons have like charge half the
time.

For $\mu = 4\:M_W$, the major contributions are from
\be
\label{44}
\tilde g \rightarrow \bar qq' \chi^\pm_1,\: \bar tb\chi^+_1 + h.c., \:
\bar tb \chi^+_2 + h.c.
\ee
followed by the leptonic decays of \eq{37}, \eq{38} and \eq{41}.  The
corresponding LSD signal is shown in Fig. 3.  The size of this signal
is a little less than half of that for $\mu = -M_W$.  The reason of
course is the reduced top quark contribution to gluino decay, as
remarked earlier.  Fig 3 also shows the signal for $\mu = 6\:M_W$.  In
this case the last channel of \eq{44} is kinematically forbidden.
Consequently there is no dependence on the charged Higgs mass.  There
is only a slight reduction in the signal in going from $\mu = 4M_W$ to
$6M_W$.  The signal is expected to remain at this level for higher
values of $\mu$ as well.

The LSD signals for the 800 GeV gluino, shown in Figs. 2 and 3, can be
separated from the LSD background of \eq{35}, shown in Fig. 1, by a
$p_{T2} > 50$ GeV cut.  But they remain below the level of the fake
LSD background from \eq{36}.  However, they can be distinguished by
the amount of missing-$p_T$ accompanying the LSD events.  Fig. 4 shows
the LSD signals for 300 and
800 GeV gluino along with both the backgrounds against the
accompanying missing-$p_T$.  A $p_T$ cut of 20 GeV has been applied on
both the leptons.  There is a significantly larger amount of
missing-$p_T$ accompanying the LSD signal compared to either
background.  The reason of course is that the signal events are
accompanied by a pair of LSPs in additionto the neutrinos, as
remarked earlier.  Fig. 4 shows the LSD signals, only for $\mu =
4M_W$, since the corresponding signals for $\mu = -M_W$ are very
similar for $M_{\tilde g} = 300~{\rm GeV}$ and larger for $M_{\tilde
g} = 800~{\rm GeV}$.

Finally, Fig. 4 shows that the LSD signal for both the gluino masses
can be separated from the backgrounds by an accompanying missing-$p_T$
cut that retains about half the signal size.  For a 800 GeV gluino,
the size of the surviving signal is $\sim 5\: fb$, corresponding to
$\sim 50$ events per year for the typical low luminosity $(\sim 10\:
fb/{\rm year})$ option of LHC.  Thus one expects a viable LSD signal
for a 800 GeV gluino over practically the full parameter space of MSSM
even at the low luminosity option of LHC.  The signal goes down by a
factor of 3-4 for $M_{\tilde g} = 1000~{\rm GeV}$ and $\sim 10$ for
$M_{\tilde g} = 1200~{\rm GeV}$.  Thus the LSD signal is exepcted to
remain viable upto a gluino mass of 1200 GeV at the high luminosity
option of LHC, with an expected lumninosity of $\sim 100\: fb/{\rm
year}$.

\section{Summary}

We have undertaken a systematic analysis of the LSD signature for
gluino production at LHC in the $R$-conserving minimal supersymemtric
standard model, taking into account the top quark and Higgs boson
effects in the cascade decay.  We have considered two representative
values of gluino mass, 300 and 800 GeV, along with those of the other
SUSY parameters --- $\mu, \tan \beta$ and $M_{H^\pm}$.  The top quark
mass has been taken to be 175 GeV, as suggested by the recent CDF data
[24].  The main results are summarized below.

For a relatively low gluino mass of $\sim 300$ GeV, the top quark
contribution is kinematically suppressed.  Here the main contribution
to the LSD signal comes from the three-body decay of the lighter
chargino $(\chi^\pm_1)$ into the LSP $(\chi^0_1)$ via a virtual $W$
boson.  The signal is seen to be insensitive to $\mu, \tan \beta$ as
well as $M_{H^\pm}$.  For a large gluino mas of $\sim 800$ GeV, the
top quark contribution to the LSD signal is very important,
particularly at small $\mu$.  Consequently one gets the largest signal
at small $\mu$; but it remains viable at larger values of $\mu$ as
well.  It is insensitive to the other parameters.  A suitable cut on
the accompanying missing-$p_T$ can separate the gluino signal from the
underlying LSD background, while retaining about half the signal size.
Even in the unfavourable case of large $\mu$, the size of the
surviving signal for a 800 GeV gluino is $\sim 5\: fb$.  This
corresponds to $\sim 50$ events for the low luminosity opiton of LHC.
Thus the LSD signal provides an unambiguous signature for gluino
production upto $\sim$ 800 GeV at the low luminosity option of LHC.
At the high luminosity option the signature remains viable upto a
gluino mass of $\sim$ 1200 GeV.

We are grateful to the organisers of WHEPP-3 at Madras last January,
where this investigation was started as a working group project.  We
are also grateful to our colleagues from far and near, Mike Bisset,
Manual Drees, Rohini Godbole, N.K. Mondal, P.N. Pandita, Probir Roy
and Xerxes Tata for discussions.  One of us (M.G.) acknowledges
financial support from the Council of Scientific and Industrial
Research, India.

\newpage

\noindent \underbar{\bf References}

\begin{enumerate}

\item Report of the Supersymmetry working Group (C. Albajar et al),
Proc. of ECFA - LHC Workshop, Vol. II, p.606 -- 683, CERN 90-10
(1990).

\item For a review see e.g. H. Haber and G. Kane, Phys. Rev.
\underbar{117}, 75 (1985).

\item J. Ellis, J. Hagelin, D. Nanopoulos, K. Olive and M. Srednicki,
Nucl.  Phys. \underbar{B238}, 453 (1984).

\item H. Baer, V. Barger, D. Karatas and X. Tata, Phys. Rev.
\underbar{D36}, 96 (1987).

\item M. Barnett, J. Gunion and H. Habber, Phys. Rev. \underbar{D37},
1892 (1988); Proc. of 1988 Summer Study on High Energy Physics, Snowmass,
Colorado (World Scientific, 1989) p.230; Proc. of 1990 Summer Study on
High Energy Physics, Snowmass, Colorado (World Scientific, 1992)
p.201; Phys. Lett. \underbar{B315}, 349 (1993).

\item H. Baer, X. Tata and J. Woodside, Phys. Rev. \underbar{D41}, 906
(1990).

\item H. Baer, C. Kao and X. Tata, Phys. Rev. \underbar{D48}, R2978
(1993).

\item D.P. Roy, Phys. Lett. \underbar{B196}, 395 (1987).

\item H. Dreiner, M. Guchait and D.P. Roy, Phys. Rev. \underbar{D49},
3270 (1994).

\item N.K. Mondal and D.P. Roy, Phys. Rev. \underbar{D49}, 183 (1994).

\item A. Bartl, W. Majerotto, B. Mosslacher, N. Oshimo and S. Stippel,
Phys. Rev. \underbar{D 43}, 2214 (1991).

\item H. Baer, X. Tata and J. Woodside, Phys. Rev. \underbar{D45}, 142
(1992); H. Baer, M. Bisset, X. Tata and J. Woodside, Phys. Rev.
\underbar{D46}, 303 (1992); A. Bartl, W. Majerotto, B. Mosslacher and
N. Oshimo, Z. Phys. \underbar{C52}, 477 (1991).

\item J. Gunion and H. Habber, Phys. Rev. \underbar{D37}, 2515 (1988).

\item H. Baer, A. Bartl, D. Karatas, W. Majerotto and X. Tata, Int.
Journal of Mod. Phys. \underbar{A4}, 4111 (1989).

\item H. Baer, X. Tata and J. Woodside, Phys. Rev. \underbar{D42},
1568 (1990).

\item A.D. Martin, R.G. Roberts and W.J. Sterling, Phys. Lett.
\underbar{B306}, 145 (1993);
\underbar{B309}, 492 (1993).

\item M. Gluck, E. Hoffman and E. Reya, Z. Phys. \underbar{C13}, 119
(1982).

\item M. Gluck, E. Reya and A. Vogt, Z. Phys. \underbar{C48}, 471
(1990); Phys. Lett. \underbar{B306}, 391 (1993).

\item This also applied to the LSD signal of [9] for the $R$-violating
SUSY model, while the top quark and Higgs boson effects are
insignificant in this case.

\item J. Gunion and H. Habber, Nucl. Phys. \underbar{B272}, 1 (1986).

\item M. Guchait, Z. Phys. \underbar{C57}, 157 (1993).

\item L. Roszkowski, Phys. Lett. \underbar{B262}, 59 (1991).

\item Review of Particle Properties, Phys. Rev. \underbar{D50}, 1173
-- 1826 (1994).

\item CDF Collaboration: F. Abe et al., Phys. Rev. Lett.
\underbar{73}, 225 (1994); Fermilab-Pub-94/097-E (Submitted to Phys.
Rev. D)

\item Although there are marginal regions of the parameter space for
which $\chi^0_2 \rightarrow \chi^0_1 H^0_{2,3}$ two-body decays are
not yet strictly ruled out.

\item There is however a small region of the parameter space,
corresponding to a small $M_{H^\pm} (\sim 100~{\rm GeV})$ and large
$\tan \beta (\gsim 40)$ as well as $|\mu| (\gsim 500~{\rm GeV})$,
where the $H^\pm$ mediated decay \eq{27} may dominate over the $W$
mediated decay \eq{25}.  The first two conditions ensure that the
suppressions from the $H^\pm$ propagator and decay vertex are not very
large.  Moreover the last two conditions imply $G_{L,R} \ll F_{L,R}$,
since $N_{12} \ll N_{13,14}$ and $U_{12} \ll U_{11} (V_{12} \ll
V_{11})$ in this region [13].  Consequently there may not be a viable
LSD signal in this corner of the parameter space.

\item There is also a LSD background from $b\bar b$ via missing.  But
it is strongly suppressed by the isolation cut on the leptons [10].
Morever there is very little mixing-$P_T$ accompanying this
background.  Several other sources of LSD background have been
discussed by Baer et al [12].

\end{enumerate}

\newpage

\noindent \underbar{\bf Figure Captions}

\begin{description}

\item[\rm Fig. 1.] The LSD signals for 300 GeV gluino production at
LHC shown against the $p_T$ of the 2nd (softer) lepton for $\mu = \pm
4M_W, -M_W$ and $\tan \beta = 2,10$.  Also shown are the LSD
backgrounds from $t\bar t$ production --- a real background arising
from the leptonic decay of one $t$ via $b$ (crosses) and a fake
background arising from the misidentification of one of the lepton
charges (dots).

\item[\rm Fig. 2.]  The LSD signals for 800 GeV gluino production at
LHC for $\mu = -M_W$, $M_{H^\pm} = 500, 100$~ GeV and $\tan \beta = 2,
10$.

\item[\rm Fig. 3.] The LSD signals for 800 GeV gluino production at
LHC for $\mu = 4M_W, M_{H^+} = 500, 100$ GeV and $\tan \beta = 2,10$.
Also shown are the signals for $\mu = 6M_W$ which are practically
independent of $M_{H^\pm}$.

\item[\rm Fig. 4.] The accompanying missing-$p_T$ $(p\!\!\!/_T)$
distribution of the LSD signals for 300 and 800 GeV gluino production
at LHC for $\mu = 4M_W, \tan \beta = 10$ and $M_{H^\pm} = 500$~ GeV.
The real and fake LSD backgrounds from $t \bar t$ production are shown
by long-dashed and dotted lines respectively.

\end{description}

\newpage

\begin{center}
{\sf Table I.  Masses (in GeV) and gluonic branching fractions
of the chargino and neutralino states for $M_{\tilde g}= 300~{\rm
GeV}$}
\end{center}

\smallskip
$$
\begin{tabular}{|cccccc|}
\hline
$\tan \beta$ & $\mu$ & $M_{\chi^\pm_i}$ & $B_{\tilde g \rightarrow
\bar qq'\chi^\pm_i}$ & $M^{\chi^0_i}$ & $B_{\tilde g \rightarrow
\bar q q \chi^0_i}$ \\
\hline
&&&&& \\
& $4M_W$ & 77.7 & .47 & 42.3 & .20 \\
&&&& 81.8 & .32 \\
&& 349.8 & 0 & -326.4 & 0 \\
&&&& 352.9 & 0 \\
&&&&& \\
& $-4M_W$ & 110.6 & .47 & 53.6 & .20 \\
&&&& 110.7 & .32 \\
2 & & 340.8 & 0 & 328.0 & 0 \\
&&&& -341.9 & 0 \\
&&&&& \\
& $-M_W$ & 91.5 & .30 & 54.9 & .20 \\
&&&& 73.1 & .11 \\
&& 146.2 & .19 & $-118.4$ & .04 \\
&&&& 141.3 & .16 \\
&&&&& \\
& $M_W$ & 16.6 && $-.01$ & \\
&&&& 63.3 & \\
&& 171.6 && $-87.5$ & \\
&&&& 175.0 & \\
&&&&& \\
\hline
&&&&& \\
& $4M_W$ & 89.8 & .52 & 48.2 & .17 \\
&&&& 90.6 & .31  \\
&& 346.9 & 0 & $-332.5$ & 0 \\
&&&& 344.3 & 0 \\
&&&&& \\
& $-4M_W$ & 97.9 & .50 & 50.9 & .18 \\
&&&& 97.8 & .31 \\
&& 344.7 & 0 & $-336.3$ & 0 \\
10 &&&& 338.2 & 0 \\
&&&&& \\
& $-M_W$ & 58.0 & .42 & 37.2 & .15 \\
&&&& 65.0 & .22 \\
&& 162.4 & .10 & $-109.0$ & .04 \\
&&&& 157.5 & .08 \\
&&&&& \\
& $M_W$ & 40.6 && 23.3 & \\
&&&& 63.9 & \\
&& 167.6 && -101.8 & \\
&&&&165.3& \\
\hline
\end{tabular}
$$

\newpage

\begin{center}
{\sf Table II.  Masses (in GeV) and gluonic branching fractions
of chargino and neutralino states for $M_{\tilde g} = 800~{\rm GeV}$}
\end{center}

\smallskip

$$
\begin{tabular}{|cccccccc|}
\hline
$\tan \beta$ & $\mu$ & $M_{\chi^\pm_i}$ & $B_{\tilde g \rightarrow \bar
qq' \chi^\pm_i}$ & $B_{\tilde g \rightarrow \bar t b\chi^+_i + hc}$ &
$M_{\chi^0_i}$ & $B_{\tilde g \rightarrow \bar q q\chi^0_i}$ &
$B_{\tilde g \rightarrow \bar tt\chi^0_i}$ \\
\hline
&&&&&&& \\
& $-M_W$ & 92 & .02 & .32 & 75 & 0 & .11 \\
&&&&& $-103$ & 0 & .11 \\
&& 286 & .14 & .11 & 143 & .05 & .04 \\
&&&&& 286 & .07 & .03 \\
&&&&&&& \\
2 & $4M_W$ & 213 & .24 & .15 & 124 & .11 & .02 \\
&&&&& 220 & .14 & .03 \\
&& 375 & .04 & .17 & $-321$ & 0 & .04 \\
&&&&& 378 & .02 & .04 \\
&&&&&&& \\
& $6M_W$ & 243 & .39 & .16 & 127 & .15 & .03 \\
&&&&& 244 & .24 & .02 \\
&& 505 & 0 & 0 & $-481$ & 0 & 0 \\
&&&&& 508 & 0 & 0 \\
&&&&&&& \\
\hline
&&&&&&& \\
& $-M_W$ & 78 & .04 & .28 & 61 & .02 & .08 \\
&&&&& $-96$ & .01 & .12 \\
&& 291 & .12 & 13 & 146 & .05 & .04 \\
&&&&& 290 & .08 & .03 \\
&&&&&&& \\
10 & $4M_W$ & 229 & .25 & .12 & 130 & .11 & .02 \\
&&&&& 232 & .15 & .02 \\
&& 366 & .04 & .18 & $-326$ & 0 & .03 \\
&&&&& 365 & .02 & .05 \\
&&&&&&& \\
& $6M_W$ & 253 & .37 & .17 & 131 & .16 & .03 \\
&&&&& 256 & .24 & .02 \\
&& 500 & 0 & 0 & $-485$ & 0 & 0 \\
&&&&& 496 & 0 & 0 \\
&&&&&&&\\
\hline
\end{tabular}
$$

\end{document}